\documentclass[pra,twocolumn,aps,superscriptaddress,showpacs]{revtex4-1}

\usepackage{hyperref}
\usepackage{graphicx}
\usepackage{amsmath}
\usepackage{amsfonts}
\usepackage{amssymb}
\usepackage{epsfig}
\usepackage[usenames,dvipsnames]{color}
\usepackage{setspace}
\usepackage{bm}

\begin{document}
\title{Fluctuation driven topological transition of binary condensates
       in optical lattices}

\author{K. Suthar}
\affiliation{Physical Research Laboratory,
             Navrangpura, Ahmedabad-380009, Gujarat,
             India} 
\affiliation{Indian Institute of Technology,
             Gandhinagar, Ahmedabad-382424, Gujarat, India}             
\author{Arko Roy}
\affiliation{Physical Research Laboratory,
             Navrangpura, Ahmedabad-380009, Gujarat,
             India}
\affiliation{Indian Institute of Technology,
             Gandhinagar, Ahmedabad-382424, Gujarat, India}
\author{D. Angom}
\affiliation{Physical Research Laboratory,
             Navrangpura, Ahmedabad-380009, Gujarat,
             India}

\date{\today}


\begin{abstract}

We show the emergence of a third Goldstone mode in binary condensates at
the phase-separation in quasi-1D optical lattices. We develop the coupled
discrete nonlinear Schr\"odinger equations (DNLSEs) using 
Hartree-Fock-Bogoliubov theory with Popov approximation in the Bose-Hubbard 
model to investigate the mode evolution at zero temperature. In particular,
as the system is driven from miscible to immiscible phase. We demonstrate 
that the position swapping of the species in $^{87}$Rb-$^{85}$Rb system is 
accompanied by a discontinuity in the excitation spectrum. Our results show 
that in quasi-1D optical lattices, the presence of the fluctuations 
dramatically change the geometry of the ground state density profile of TBEC. 

\end{abstract}

\pacs{42.50.Lc, 67.85.Bc, 67.85.Fg, 67.85.Hj}


\maketitle

\section{Introduction}

 Ultracold dilute atomic Bose gases in low dimensions have been the subject of 
growing interest over the last few decades. These are an ideal platform to
probe many-body phenomena where quantum fluctuations play a crucial 
role~\cite{cazalilla_11,lieb_63}. In particular, the use of optical lattices 
serve as an excellent and versatile tool to study the physics of 
strongly correlated systems, and other phenomena in condensed matter 
physics~\cite{morsch_06,bloch_08}. A variety of experimental techniques have 
been used to load and manipulate Bose-Einstein condensates (BECs) in optical 
lattices~\cite{friebel_98,guidoni_98,belen_04,bloch_05}. These have helped to 
explore quantum phase transition~\cite{sachdev_11} namely 
superfluid (SF)--Mott insulator (MI) 
transition~\cite{greiner_02,moritz_03,stoferle_04,tolra_04}. 
The characteristics of SF phase, such as coherence~\cite{orzel_01,greiner_01}, 
collective modes~\cite{fort_03} and transport~\cite{fertig_05,fallani_04} 
have also been observed. The center of mass dipole oscillation of BEC in a 
cigar-shaped lattice potential has been experimentally studied in 
detail~\cite{burger_01}. In such systems, a decrease in the Kohn mode frequency 
has been reported in Ref.~\cite{cataliotti_01} which has been justified in 
Ref.~\cite{kramer_03} as the increase of the effective mass due to the lattice 
potential. On the theoretical front, the low-lying collective excitations of a 
trapped Bose gas in periodic lattice potential have been studied in 
Refs.~\cite{lundh_04,martikainen_03,rey_04,amrey_04} using Bose-Hubbard (BH) 
model~\cite{fisher_89}. 

 The two-component BEC (TBEC), on the other hand, exhibits an unique property
that they can be phase-separated~\cite{navarro_09}. There have been numerous 
experimental and theoretical investigations of binary mixtures of BECs over 
the last few years. Experimentally, it is possible to vary the interactions 
through Feshbach resonance~\cite{papp_08,tojo_10}, and drive the binary mixture 
from miscible to immiscible phase or vice-versa. Among the various lines of 
investigation, the theoretical study of the stationary states~\cite{gautam_11}, 
dynamical instabilities~\cite{gautam_10,kadokura_12} and the collective 
excitations~\cite{ticknor_13,gordon_98} of TBECs are noteworthy. Furthermore, 
in optical lattices TBECs have also been observed in recent 
experiments~\cite{catani_08,gadway_10}.
 Theoretical studies of 
TBECs in optical lattices~\cite{chen_03,kuklov_03,kuklov_04,cha_13} and, in 
particular, phase-separation~\cite{zhan_14,mishra_07,kuo_08} and dynamical 
instabilities~\cite{ruostekoski_07} have also been carried out. Despite all 
these theoretical and experimental advancements, the study of collective 
excitations of TBECs in optical lattices is yet to be explored. This is the 
research gap addressed in the present work. 

 In this paper, we report the development of coupled discrete nonlinear 
Schr\"odinger equations (DNLSEs) of TBECs in optical lattices under 
Hartree-Fock-Bogoliubov-Popov approximation~\cite{griffin_96}. We use this 
theory to study the ground state density-profiles and the quasiparticle
spectrum of $^{87}$Rb-$^{85}$Rb and $^{133}$Cs-$^{87}$Rb TBECs at zero 
temperature. We, in particular, focus on the evolution of the quasiparticle 
as the TBEC is driven from miscible to immiscible phase. This is possible by 
the tuning either the intra- or interspecies interaction strengths. The two 
systems considered correspond to these possibilities. The fluctuation and 
interaction induced effect on the collective excitation spectra and topological change in density profiles is the major finding of our present study. It 
deserves to be mentioned here that for systems without the lattice potential, 
at equilibrium, recent works have shown the existence of additional Goldstone 
modes in TBECs at phase-separation~\cite{arko_14} and complex eigenenergies 
due to quantum fluctuations~\cite{arko_14a}.

 The paper is organized as follows. Sec.~\ref{1D_ol} describes the 
tight-binding approximation for a trapped BEC in 1D lattice potential. In 
Sec.~\ref{hfb_popov} we present the HFB-Popov theory to determine the
quasiparticle energies and mode functions of single component BEC and TBECs 
at finite temperature. The results of our studies are presented in 
Sec.~\ref{results}. 
Finally, we highlight the key results of our work in Sec.~\ref{conc}.


\section{Quasi-1D optical lattice}
\label{1D_ol}

We consider a Bose-Einstein condensate (BEC), held within a highly anisotropic
cigar shaped harmonic potential with trapping frequencies 
$\omega_x = \omega_y = \omega_{\perp} \gg \omega_z$. In this case we can
integrate out the condensate wave-function along $x$ and $y$-direction and 
reduce it to a quasi-1D condensate. In the mean-field approximation, the 
grand-canonical Hamiltonian, in the second quantized form, of the 
bosonic atoms in an external potential plus lattice is given by
\begin{eqnarray}
\hat{H} = &&\int dz \hat{\Psi}^{\dagger}(z)\left(-\frac{\hbar^2}{2m}
             \frac{\partial^2}{\partial z^2} 
        + V_{\rm latt}(z)\right)\hat{\Psi}(z) \nonumber \\
      &+& \int dz (V_{\rm ext} - \mu) \hat{\Psi}^{\dagger}(z) \hat{\Psi}(z)
          \nonumber\\
      &+& \frac{1}{2} \int dz dz'\hat{\Psi}^{\dagger}(z)
          \hat{\Psi}^{\dagger}(z') U(z-z') \hat{\Psi}(z)\hat{\Psi}(z'),
\end{eqnarray}
where $\hat{\Psi}(z)$ and $\hat{\Psi}^{\dagger}(z)$ are the bosonic field
operators which obey the Bose commutation relations, $m$ is the atomic mass
of the species, $V_{\rm latt}$ is the periodic lattice potential, 
$V_{\rm ext}$ is the external trapping potential, $\mu$ is the chemical 
potential and $U = 2\sqrt{\lambda\kappa}\hbar^2 N a_s/m$, with $N$ as the
total number of atoms, $\lambda = \omega_x/\omega_z$ and 
$\kappa = \omega_y/\omega_z$ are the anisotropy parameters along $x$- and
$y$-direction, and $a_s$ as the $s$-wave scattering length, which is repulsive 
($a_s > 0$) in the present work. The net external potential is 
\begin{eqnarray}
 V &=& V_{\rm ext} + V_{\rm latt} \nonumber \\
   &=& \frac{1}{2} m\omega^2_{z} z^2 + V_0 \sin^2(kz),
\end{eqnarray}
where $V_0 = s E_R$ is the optical lattice depth with $s$ and $E_R$ as 
the lattice depth scaling parameter, and the recoil energy of the laser light 
photon, respectively. The wave number of the counter-propagating laser
beams, which are used to create periodic lattice potential is $k = \pi/a$ with 
$a = \lambda_L/2$ is the lattice spacing and $\lambda_L$ is the wavelength of 
the laser light. The energy barrier between adjacent lattice sites is 
expressed in units of the recoil energy $E_R = \hbar^2 k^2/2m$. 
In tight binding approximation, valid when $\mu \ll V_0$, the 1D field 
operator can be written as~\cite{chiofalo_00} 
\begin{equation}
 \hat{\Psi}(z) =  \sum_{j} \hat{a}_{j} \phi_{j}(z),
\end{equation}
where $\hat{a}_{j}$ is the annihilation operator corresponding to the 
$j$th site, and the spatial part $\phi_{j}(z) = \phi(z-ja)$ is the 
orthonormal Gaussian orbital of the lowest vibrational band centered at the 
$j$th lattice site, with 
$\smallint dz~\phi^{*}_{j \pm 1}(z)\phi_{j}(z) = 0$ and 
$\smallint dz~|\phi_{j}(z)|^2 = 1$. By using above ansatz
in $\hat{H}$ and considering only the nearest neighbour tunneling we
obtain the Bose-Hubbard (BH) Hamiltonian.


\section{HFB-Popov approximation}
\label{hfb_popov}

\subsection{Single-component BEC in optical lattices}
\label{hfb_popov_1s}
 The BH Hamiltonian describes the dynamics of 1D optical lattices when only 
the lowest band or the lowest vibrational level of the site is occupied. 
In this case the tight binding approximation~\cite{jaksh_98} is valid, and the 
BH Hamiltonian of the system is
\begin{equation}
 \hat{H} = -J\sum_{\langle jj'\rangle} \hat{a}^{\dagger}_{j} \hat{a}_{j'}
           + \sum_j \left[(\epsilon_j - \mu) \hat{a}^{\dagger}_j\hat{a}_{j} 
           + \frac{1}{2} U \hat{a}^{\dagger}_j\hat{a}^{\dagger}_j
             \hat{a}_{j}\hat{a}_{j}\right],
\end{equation}
where the index $j$ runs over the lattice sites, $\langle jj'\rangle$
represents the nearest neighbour sum, and $\hat{a}_j(\hat{a}^{\dagger}_j)$ is 
the bosonic annihilation (creation) operator of a bosonic atom at the $j$th
lattice site. 
Here $J = \smallint dz~\phi_{j+1}^*(z)[-(\hbar^2/2m)(\partial^2/\partial z^2) 
+ V_0 \sin^2(2\pi z/\lambda_L)]\phi_{j}(z)$ is the tunneling matrix
element between adjacent sites, $\epsilon_j = \smallint dz~V_{\rm ext}(z)
|\phi_{j}(z)|^2$ is the energy offset of the $j$th lattice site, and
$U = (2\sqrt{\lambda\kappa}\hbar^2 N a_s/m) \smallint dz |\phi_{j}(z)|^4$
is the on-site interaction strength of atoms occupying the $j$th lattice
site. The offset energy can also be expressed as $\epsilon_j = j^2 \Omega$,
here, $\Omega = m\omega^2_{z} a^2/2$ is the energy cost to move a boson
from the central site to its nearest neighbour site. To take into account the 
quantum fluctuations and thermal effects in the description of the system, we 
decompose the Bose field operator of each lattice site $j$ in terms of a 
complex mean-field part $c_j$ and a fluctuation operator 
$\hat{\varphi}_j$, as $\hat{a}_j = (c_j +\hat{\varphi}_j)e^{-i\mu t/\hbar}$.
Using this field operator in the BH Hamiltonian, we get
\begin{equation}
 \hat{H} = H_0 + \hat{H}_1 + \hat{H}_2 + \hat{H}_3 + \hat{H}_4,
\end{equation}
with 
\begin{subequations}
 \begin{eqnarray}
  H_0  =  &-&J \sum_{\langle jj'\rangle} c^{*}_j c_{j'} 
          + \sum_j \left[(\epsilon_j-\mu) |c_j|^2 
          + \frac{1}{2} U |c_j|^4 \right],~~~~~~~~\\
  \hat{H}_1 = &-&J \sum_{\langle jj'\rangle} \hat{\varphi}_j c^{*}_{j'}
                +  \sum_j \left(\epsilon_j - \mu + U|c_j|^2 \right) c^{*}_j
                   \hat{\varphi}_j  + \rm{h.c.}, \nonumber\\~\\
  \hat{H}_2 = &-&J \sum_{\langle jj'\rangle} \hat{\varphi}^{\dagger}_j 
                   \hat{\varphi}_{j'} + \sum_j (\epsilon_j - \mu)
                   \hat{\varphi}^{\dagger}_j\hat{\varphi}_j \nonumber\\   
               &+& \frac{U}{2} \sum_j \left(\hat{\varphi}^{\dagger 2}_j c^2_j 
                +  \hat{\varphi}^{2}_j c^{*2}_j 
                + 4|c_j|^2 \hat{\varphi}^{\dagger}_j\hat{\varphi}_j \right), \\
  \hat{H}_3 =&&U  \sum_j \hat{\varphi}^{\dagger}_j\hat{\varphi}^{\dagger}_j
                   \hat{\varphi}_j c_j + \rm{h.c.}, \\
  \hat{H}_4 =&&\frac{U}{2} \sum_j \hat{\varphi}^{\dagger}_j
                \hat{\varphi}^{\dagger}_j \hat{\varphi}_j\hat{\varphi}_j,
 \end{eqnarray}
 \label{ham}
\end{subequations}
where subscript of the various terms indicates the order of fluctuation
operators and $\rm{h.c.}$ stands for the hermitian conjugate. To 
study the system without quantum fluctuation at $T=0 K$, we consider terms up 
to second order in $\hat{\varphi}_j$ and neglect the higher order terms 
(third and fourth order). The lowest order term of the Hamiltonian describes 
the condensate part of the system. The minimization of $H_0$ with respect to 
the variation in the complex amplitude $c^{*}_j$ gives the time independent 
DNLSE, which can be written as
\begin{equation}
 \mu c_j = -J(c_{j-1} + c_{j+1}) + (\epsilon_j + U n^c_j) c_j, 
\end{equation}
with the condensate density $n^c_j = |c_j|^2$. The quadratic Hamiltonian
$\hat{H}_2$ is the leading order term which describes the noncondensate 
part, since the variation in $\hat{H}_1$  vanishes by using the fact that 
$c_j$ is a stationary solution of the DNLSE. The minimization of $\hat{H}_2$ 
yields the governing equation for the noncondensate given by
\begin{equation}
 \mu \hat\varphi_j = -J(\hat\varphi_{j-1} + \hat\varphi_{j+1}) + (\epsilon_j 
                  + 2 U n^c_j)\hat\varphi_j + U c^2_j \hat\varphi^{\dagger}_j.
 \label{non_cond}
\end{equation}
The quadratic Hamiltonian can be diagonalized using the Bogoliubov 
transformation
\begin{subequations}
 \begin{eqnarray}
  \hat\varphi_j &=& \sum_l\left[u^l_j\hat{\alpha}_l e^{-i\omega_l t} 
                      - v^{*l}_j\hat{\alpha}^{\dagger}_l e^{i\omega_l t}\right],\\
  \hat\varphi^{\dagger}_j &=& \sum_l\left[u^{*l}_j\hat{\alpha}^{\dagger}_l 
              e^{i\omega_l t} - v^{l}_j\hat{\alpha}_l e^{-i\omega_l t}\right],
 \end{eqnarray}
 \label{bog_trans}                        
\end{subequations}
where $u^l_j$ and $v^l_j$ are the quasiparticle amplitudes, 
$\omega_l=E_l/\hbar$ is the $l$th quasiparticle mode frequency with $E_l$ 
as the mode energy, and $\hat{\alpha}_l(\hat{\alpha}^{\dagger}_l)$ are the 
quasiparticle annihilation (creation) operators, which satisfy the Bose 
commutation relations. The quasiparticle amplitudes satisfy the following 
normalization conditions
\begin{subequations}
 \begin{eqnarray}
  \sum_j \left(u^{*l}_j u^{l'}_j - v^{*l}_j v^{l'}_j\right) &=& \delta_{ll'},\\
  \sum_j \left(u^l_j v^{l'}_j - v^{*l}_j u^{*l'}_j\right) &=& 0.
 \end{eqnarray}
\end{subequations}
By using the definitions of $\hat\varphi_j$ from Eq.~(\ref{bog_trans}) in 
$\hat{H}_2$ [Eq.~(\ref{ham}c)] and using the above conditions, we get the 
following Bogoliubov-de Gennes (BdG) equations
\begin{subequations}
 \begin{eqnarray} 
  E_l u^l_j &=& -J(u^l_{j-1} + u^l_{j+1}) + [2 U n^c_{j} 
                + (\epsilon_j - \mu)]u^l_j - U c^2_j v^l_j, \nonumber\\~\\
  E_l v^l_j &=& J(v^l_{j-1} + v^l_{j+1}) - [2 U n^c_{j} 
               + (\epsilon_j -\mu)]v^l_j + U c^{*2}_j u^l_j. \nonumber\\
 \end{eqnarray}
 \label{bdg_eq}              
\end{subequations}
This set of coupled equations describe the quasiparticles of condensate in 
the optical lattice without considering the quantum fluctuations.

 To investigate the effect of fluctuation and finite temperature we include 
the higher order terms ($\hat{H}_3$ and $\hat{H}_4$) of the fluctuation 
operator in the Hamiltonian. We treat these terms in the  self-consistent 
mean-field approximation~\cite{griffin_96} such that $\hat{\varphi}^{\dagger}_j 
\hat{\varphi}_j \hat{\varphi}_j \approx 2\tilde{n}_j \hat{\varphi}_j 
+ \tilde{m}_j \hat{\varphi}^{\dagger}_j$ and $\hat{\varphi}^{\dagger}_j 
\hat{\varphi}^{\dagger}_j \hat{\varphi}_j \hat{\varphi}_j 
\approx 4\tilde{n}_j\hat{\varphi}^{\dagger}_j\hat{\varphi}_j + \tilde{m}_j 
 \hat{\varphi}^{\dagger}_j \hat{\varphi}^{\dagger}_j + \tilde{m}^{*}_j
 \hat{\varphi}_j\hat{\varphi}_j - (2\tilde{n}^2_j + |\tilde{m}_j|^2)$, 
 where $\tilde{n}_j = \langle\hat{\varphi}^{\dagger}_j
 \hat{\varphi}_j\rangle$ and $\tilde{m}_j = \langle\hat{\varphi}_j
 \hat{\varphi}_j\rangle$ are the excited population (noncondensate) density 
and anomalous density at the $j$th site, respectively. In the HFB-Popov 
approximation, where the anomalous density is neglected, the corrections from 
higher order terms yield the modified DNLSE
\begin{equation}
 \mu' c_j = -J(c_{j-1} + c_{j+1}) + [\epsilon_j + U (n^c_j + 2\tilde{n}_j)]c_j,
 \label{hfb_sol}
\end{equation}
where $\mu'$ is the modified chemical potential.
The total density is $n = \sum_j (n^c_j + \tilde{n}_j)$. The diagonalization 
of the modified Hamiltonian leads to the following HFB-Popov equations
\begin{subequations}
 \begin{eqnarray}
   E_l u^l_j = &-& J(u^l_{j-1} + u^l_{j+1}) + [2 U (n^c_{j} + \tilde{n}_j)
                 + (\epsilon_j - \mu')]u^l_j \nonumber\\ 
               &-& U c^2_j v^l_j, \\
   E_l v^l_j = & & J(v^l_{j-1} + v^l_{j+1}) - [2 U (n^c_{j} + \tilde{n}_j)
                 + (\epsilon_j - \mu')]v^l_j \nonumber\\
               &+& U c^{*2}_j u^l_j,
 \end{eqnarray}
 \label{hfb_eq_1s}
\end{subequations}
with the noncondensate density at the $j$th lattice site given by
\begin{equation}
 \tilde{n}_j = \sum_l [(|u^l_j|^2 + |v^l_j|^2)N_0(E_l) + |v^l_j|^2], 
\end{equation}
where $N_0(E_l) = \langle\hat{\alpha}^{\dagger}_l \hat{\alpha}_l\rangle
                = (e^{\beta E_l} - 1)^{-1}$ is the Bose-Einstein distribution 
function of the quasiparticle state with real and positive mode energy $E_l$. 
The coupled Eqs.~(\ref{hfb_sol}) and (\ref{hfb_eq_1s}) are solved iteratively 
until the solutions converge to desired accuracy. It is important to note that, at $T = 0 K$, $N_0(E_l)$ in the above equation vanishes. The noncondensate 
density, then, has contribution from only the quantum fluctuations, which is 
given by
\begin{equation}
 \tilde{n}_j = \sum_l |v^l_j|^2. 
\end{equation}
Therefore, we solve the equations self-consistently in the presence of the 
quantum fluctuations.


\subsection{Two-component BEC in optical lattices}
\label{hfb_popov_2s}
For two species condensate, the 1D second quantized grand canonical Hamiltonian
is given by
\begin{eqnarray}
 \hat{H}& = &\sum_{i=1}^{2} \int dz \hat{\Psi}^{\dagger}_{i}(z)
             \bigg [-\frac{\hbar^2}{2 m_i} \frac{\partial^2}{\partial z^2} 
             + V^{i}(z) - \mu_i + \frac{U_{ii}}{2}\hat{\Psi}^{\dagger}_{i}(z) 
                              \nonumber \\
           &\times&   \hat{\Psi}_{i}(z) \bigg ]\hat{\Psi}_{i}(z)
            + U_{12} \int dz \hat{\Psi}^{\dagger}_{1}(z)
            \hat{\Psi}^{\dagger}_{2}(z)\hat{\Psi}_{1}(z)\hat{\Psi}_{2}(z),
            \nonumber\\
 \label{ham_2sp}              
\end{eqnarray}
where $i = 1,2$ denotes the species index, $\hat{\Psi}_i$'s are the 
annihilation field operators for two different species, $\mu_i$ is the chemical 
potential of the $i$th species, $U_{ii}$ are the intraspecies interaction 
parameters, and $U_{12}$ is the interspecies interaction parameter with 
$m_i$'s as the atomic masses of the species. Here, we consider repulsive
interactions, $U_{ii},U_{12}>0$. The external potential $V^{i}$ is the sum of 
harmonic and periodic optical lattice potential. It is given by
\begin{eqnarray}
 V^{i} &=& V^{i}_{\rm ext} + V^{i}_{\rm latt} \nonumber \\
    &=& \frac{1}{2} m_i \omega_{z_i}^2 z_{i}^2 + V_0 \sin^2(2\pi z_i/\lambda_L).
\end{eqnarray}
In the present work, we consider the same external potential for 
both the species. The depth of the lattice potential is also same for both 
species which is $V_0 = sE_R$ with $E_R = \hbar^2 k^2/2m_1$. If the lattice is 
deep enough, the tight-binding approximation is valid, and the bosons can be 
assumed to occupy the lowest vibrational band only. Under this approximation, 
the Bose field operator for the two species can be expanded as
\begin{equation}
 \hat{\Psi}_{i}(z) =  \sum_{j} \hat{a}_{ij} \phi_{ij}(z),
 \label{psi}
\end{equation}
where $\hat{a}_{ij}$'s are the annihilation operators and $\phi_{ij}(z)$'s are 
the orthonormal Gaussian basis of the two species. For the present work we 
assume that the width of the Gaussian basis are identical for both the 
species ($\phi_{1j}=\phi_{2j}$). The BH Hamiltonian for two species can be 
obtained by using the above ansatz in the Hamiltonian, Eq.~(\ref{ham_2sp}). 
We, then, obtain the  many-body Hamiltonian governing the system of binary 
BEC in quasi-1D optical lattice as
\begin{eqnarray}
 \hat{H} = && \sum_{i=1}^{2}\left[-\sum_{\langle jj'\rangle} J_i
         \hat{a}^{\dagger}_{ij}\hat{a}_{ij'} + \sum_j(\epsilon^{(i)}_{j}-\mu_i)
            \hat{a}^{\dagger}_{ij}\hat{a}_{ij}\right] \nonumber\\
         &+& \frac{1}{2}\sum_{i=1}^{2} U_{ii}\sum_j\hat{a}^{\dagger}_{ij}
            \hat{a}^{\dagger}_{ij}\hat{a}_{ij}\hat{a}_{ij}
          + U_{12}\sum_j \hat{a}^{\dagger}_{1j}\hat{a}_{1j}
            \hat{a}^{\dagger}_{2j}\hat{a}_{2j}.\nonumber \\
 \label{bh_2sp}           
\end{eqnarray}
Here $J_i$ are the tunneling matrix elements, and $\epsilon^{(i)}_{j}$ is the 
offset energy of species $i$ at the $j$th lattice site. In the mean-field 
approximation, using Bogoliubov approximation like in single species 
condensate we decompose the operators of both species as 
$\hat{a}_{1j} = (c_j + \hat{\varphi}_{1j})e^{-i\mu_1 t/\hbar}$ 
and $\hat{a}_{2j} = (d_j + \hat{\varphi}_{2j})e^{-i\mu_2 t/\hbar}$. We
use these definitions in the BH Hamiltonian [Eq.~(\ref{bh_2sp})] and then 
decompose the Hamiltonian into different terms according to the order of 
noncondensate operator they contain. The minimization of the lowest order 
term gives the stationary state equations or time-independent coupled DNLSEs, 
and these are given by
\begin{subequations}
 \begin{eqnarray}
  \mu_1 c_j = &-& J_1(c_{j-1} + c_{j+1}) + \left [ \epsilon^{(1)}_j 
               + U_{11} n^{c}_{1j} + U_{12} n^{c}_{2j} \right ]  c_j,
                      \nonumber \\~\\
  \mu_2 d_j = &-& J_2(d_{j-1} + d_{j+1}) + \left [ \epsilon^{(2)}_j 
               + U_{22} n^{c}_{2j} + U_{12} n^{c}_{1j} \right ] d_j, 
                      \nonumber \\
 \end{eqnarray}
 \label{dnls_2s}
\end{subequations} 
where $n^{c}_{1j} = |c_j|^2$ and  $n^{c}_{2j} = |d_j|^2$ are the condensate 
densities of the first and second species, respectively. The noncondensate 
part of the TBEC is obtained by the minimization of the quadratic Hamiltonian
\begin{subequations}
 \begin{eqnarray}
  \mu_1 \hat{\varphi}_{1j} = &-& J_1(\hat{\varphi}_{1,j-1} 
                              + \hat{\varphi}_{1,j+1})
                              + \left [ \epsilon^{(1)}_j + 2 U_{11} n^{c}_{1j}
                                \right ] \hat{\varphi}_{1j} \nonumber\\
                             &+& U_{11} c^2_j\hat{\varphi}^{\dagger}_{1j}    
                              + U_{12}(n^{c}_{2j} \hat{\varphi}_{1j} 
                              + d^{*}_j c_j \hat{\varphi}_{2j} 
                              + d_j c_j \hat{\varphi}^{\dagger}_{2j}),
                              \nonumber\\ \\
 \mu_2 \hat{\varphi}_{2j} = &-& J_2(\hat{\varphi}_{2,j-1} 
                             + \hat{\varphi}_{2,j+1})
                             + \left [ \epsilon^{(2)}_j + 2 U_{22} n^{c}_{2j}
                               \right ] \hat{\varphi}_{2j} \nonumber\\
                             &+& U_{22} d^2_j \hat{\varphi}^{\dagger}_{2j}    
                              + U_{12}(n^{c}_{1j} \hat{\varphi}_{2j} 
                              + c^{*}_j d_j \hat{\varphi}_{1j} 
                              + c_j d_j \hat{\varphi}^{\dagger}_{1j}). 
                              \nonumber \\
 \end{eqnarray}
\end{subequations}
The Bogoliubov transformation equations of the TBEC, which couples the 
positive and negative energy mode excitations, are
\begin{subequations}
 \begin{eqnarray}
  \hat\varphi_{ij} &=& \sum_l\left[u^l_{ij}\hat{\alpha}_l e^{-i \omega_l t} 
            - v^{*l}_{ij}\hat{\alpha}^{\dagger}_l e^{i \omega_l t}\right],\\
  \hat\varphi^{\dagger}_{ij} &=& \sum_l\left[u^{*l}_{ij}
                                 \hat{\alpha}^{\dagger}_l e^{i \omega_l t} 
                        - v^{l}_{ij}\hat{\alpha}_l e^{-i \omega_l t}\right],
  \label{bog_trans_2s}                        
 \end{eqnarray}
\end{subequations}
where $u^l_{ij}$ and $v^l_{ij}$ are the quasiparticle amplitudes for the
first ($i=1$) and second ($i=2$) species. The above transformation 
diagonalizes the quadratic Hamiltonian and gives the Bogoliubov-de Gennes 
(BdG) equations at $T = 0K$ for the two-component system. The inclusion 
of the higher order terms of the perturbation or fluctuation in the quadratic 
Hamiltonian gives the HFB-Popov equations for the two-component BEC
\begin{subequations}
 \begin{eqnarray}
  E_l u^l_{1,j} = &-& J_1(u^l_{1,j-1} + u^l_{1,j+1}) + \mathcal{U}_1 u^l_{1,j} 
                      - U_{11} c^2_j v^l_{1,j} \nonumber\\ 
                  &+& U_{12} c_j(d^{*}_j u^l_{2,j} - d_j v^l_{2,j}),\\
  E_l v^l_{1,j} = &~& J_1(v^l_{1,j-1} + v^l_{1,j+1}) + \underline{\mathcal{U}}_1
                       v^l_{1,j} + U_{11} c^{*2}_j u^l_{1,j} \nonumber\\
                  &-& U_{12} c^{*}_j(d_j v^l_{2,j} - d^{*}_j u^l_{2,j}),\\
  E_l u^l_{2,j} = &-& J_2(u^l_{2,j-1} + u^l_{2,j+1}) + \mathcal{U}_2 u^l_{2,j} 
                      - U_{22} d^2_j v^l_{2,j} \nonumber\\ 
                  &+& U_{12} d_j(c^{*}_j u^l_{1,j} - c_j v^l_{1,j}),\\
  E_l v^l_{2,j} = &~& J_2(v^l_{2,j-1} + v^l_{2,j+1}) + \underline{\mathcal{U}}_2
                        v^l_{2,j} + U_{22} d^{*2}_j u^l_{2,j} \nonumber\\
                  &-& U_{12} d^{*}_j(c_j v^l_{1,j} - c^{*}_j u^l_{1,j}),
 \end{eqnarray}
 \label{bdg_eq_2sp}                 
\end{subequations}
where $\mathcal{U}_1 = 2 U_{11} (n^{c}_{1j} + \tilde{n}_{1j}) 
+ U_{12} (n^{c}_{2j} + \tilde{n}_{2j}) + (\epsilon^{(1)}_j - \mu_1)$, 
$\mathcal{U}_2 = 2 U_{22} (n^{c}_{2j} + \tilde{n}_{2j}) 
+ U_{12} (n^{c}_{1j} + \tilde{n}_{1j}) + (\epsilon^{(2)}_j - \mu_2)$ with
$\underline{\mathcal{U}}_i = -\mathcal{U}_i$. The density of the noncondensate 
atoms at the $j$th lattice site is
\begin{equation}
 \tilde{n}_{ij} = \sum_l [ (|u^l_{ij}|^2 + |v^l_{ij}|^2)N_0(E_l) 
                   + |v^l_{ij}|^2],
\end{equation}
with $N_0(E_l)$ as the Bose-factor of the system with energy $E_l$
at temperature $T$. At $T = 0 K$ the noncondensate part reduces to the
quantum fluctuations
\begin{equation}
 \tilde{n}_{ij} = \sum_l |v^l_{ij}|^2.
\end{equation}
If we neglect quantum fluctuations (noncondensate part), the HFB-Popov
Eqs.~(\ref{bdg_eq_2sp}) are the BdG equations for binary BEC.

\begin{figure}[ht]
 {\includegraphics[width=8.5cm] {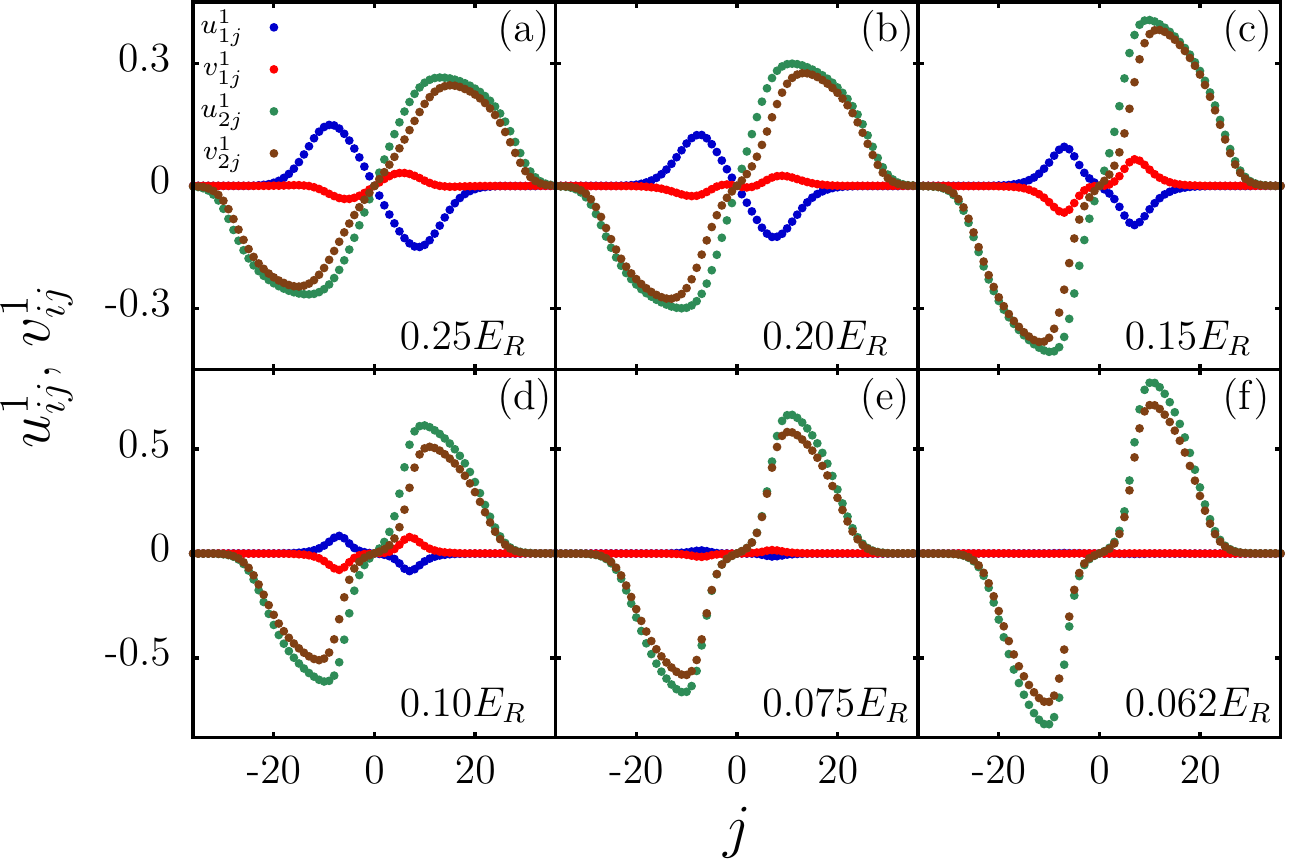}}
 \caption{The evolution of the quasiparticle amplitudes corresponding to the
          $^{85}$Rb Kohn mode as the intraspecies interaction of $^{85}$Rb
          ($U_{22}$) is decreased from $0.25 E_R$ to $0.062 E_R$. (a)-(b) When
          $U_{22}\ge 0.18 E_R$, the system is in miscible phase and the Kohn
          mode ($l =1$) have contributions from both the species, (c)-(e) when
          system is on the verge of the phase separation, then the Kohn mode of
          $^{85}$Rb goes soft, and (f) at phase separation 
          $U_{22}\le 0.065 E_R$ the Kohn mode transforms into a Goldstone 
          mode.}
 \label{mode_fns_775}
\end{figure}


\section{Results and discussions}
\label{results}

\subsection{Numerical details}
\label{num_detail}
We solve the scaled coupled DNLSE using fourth-order Runge-Kutta method to 
find the equilibrium state of the harmonically trapped binary condensates in 
optical lattices. We start the calculations for $T=0$K by ignoring the quantum 
fluctuations at each lattice site. The initial complex amplitudes of both 
species $c_j$ and $d_j$  are chosen as $1/\sqrt{N_{\rm latt}}$, with 
$N_{\rm latt}$ as the total number of lattice sites. The advantage of this
choice is that the amplitudes are normalized. We, then, use imaginary time 
propagation of the DNLSEs, Eqs.~(\ref{dnls_2s}), to find the stationary 
ground state wave function of the TBEC. In the tight binding limit, the 
condensate wave function can be defined as the superposition of the basis 
functions as shown in Eq.~(\ref{psi}). The basis function is chosen as the 
ground state, which is a Gaussian function, of lowest energy 
band~\cite{chiofalo_00}. The width of the function is a crucial parameter as 
it affects the overlap of the Gaussian orbitals at each lattice site. The 
correct estimation of the width is required in order to obtain orthonormal 
basis functions~\cite{baym_96}. Furthermore, to study the excitation spectrum, 
we cast the Eqs.~(\ref{bdg_eq_2sp}) as a matrix eigenvalue equation. The matrix 
is $4N_{\rm latt}\times4N_{\rm latt}$, non-Hermitian, non-symmetric and may 
have complex eigenvalues. To diagonalize the matrix and to find the 
quasiparticle energies $E_l$, and amplitudes $u^l_{ij}$'s and $v^l_{ij}$'s, we 
use the routine ZGEEV from the LAPACK library~\cite{anderson_99}. In the later 
part of the work, when we include the effect of the quantum fluctuations, we 
need to solve Eqs.~(\ref{dnls_2s}) and Eqs.~(\ref{bdg_eq_2sp}) 
self-consistently. For this we iterate the solution until we reach desired 
convergence in the number of condensate and noncondensate atoms. In this 
process, sometimes, we encounter severe oscillations in the number of atoms. 
To damp these oscillations and accelerate convergence we employ a successive 
over (under) relaxation technique for updating the condensate (noncondensate) 
atom densities~\cite{simula_01}. The new solutions after the iteration 
cycle (IC) are given by
\begin{subequations}
 \begin{eqnarray}
  c^{\rm new}_{j,\rm IC} = r^{\rm ov} c_{j,\rm IC} 
                                    + (1 - r^{\rm ov}) c_{j,\rm IC-1},  \\
  \tilde{n}^{\rm new}_{j,\rm IC} = r^{\rm un} \tilde{n}_{j,\rm IC} 
                                    + (1 - r^{\rm un}) \tilde{n}_{j,\rm IC-1},
 \end{eqnarray}
\end{subequations}
 where $r^{\rm ov} > 1$ ($r^{\rm un} < 1$) is the over (under) relaxation 
parameter. After the condensate and noncondensate density converge, we compute 
low-lying mode energies, and amplitude $u^l_{ij}$'s and $v^l_{ij}$'s. During 
computation, we ensure that the eigenvalues of the HFB-Popov matrix are real 
as there are no topological defects present in the system.  
\begin{figure}[ht]
 {\includegraphics[width=8.5cm] {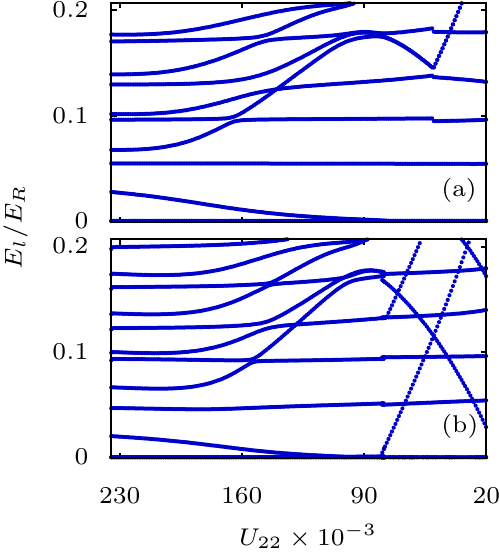}}
 \caption{The evolution of the low-lying modes as a function of the 
          intraspecies interaction of the $^{85}$Rb ($U_{22}$) in the
          $^{87}$Rb-$^{85}$Rb TBEC held in quasi-1D optical lattices. 
          (a) Excitation spectrum at zero temperature and (b) is the excitation 
          spectrum in the presence of the quantum fluctuations. Here $U_{22}$ 
          is in units of the recoil energy $E_R$.}
 \label{mode_en}
\end{figure}

\subsection{Mode evolution of trapped TBEC at $T=0K$}
\label{zero_temp}
Under the HFB-Popov approximation, the excitation spectrum of TBEC in optical
lattice is gapless for the SF phase, while it has a finite gap for the MI 
phase~\cite{greiner_02}. In SF phase, the spontaneous symmetry breaking at 
condensation results in two Goldstone modes, one each for the two species. 
The number of Goldstone modes, however, depends on whether the system is in 
miscible or immiscible phase, and geometry of the density distributions. To 
explore different possibilities, as mentioned earlier, we consider two 
different TBEC systems. These are binary mixtures which can be driven from 
miscible to immiscible phase through the variation of intra- or interspecies 
interaction using Feshbach resonance. In particular, we consider 
$^{87}$Rb - $^{85}$Rb~\cite{papp_08,handel_11} and 
$^{133}$Cs - $^{87}$Rb~\cite{mccarron_11,lercher_11} binary condensates as 
examples of the two cases, and study the mode evolution as the system 
approaches immiscible from miscible regime. 

\begin{figure}[ht]
 {\includegraphics[width=8.5cm] {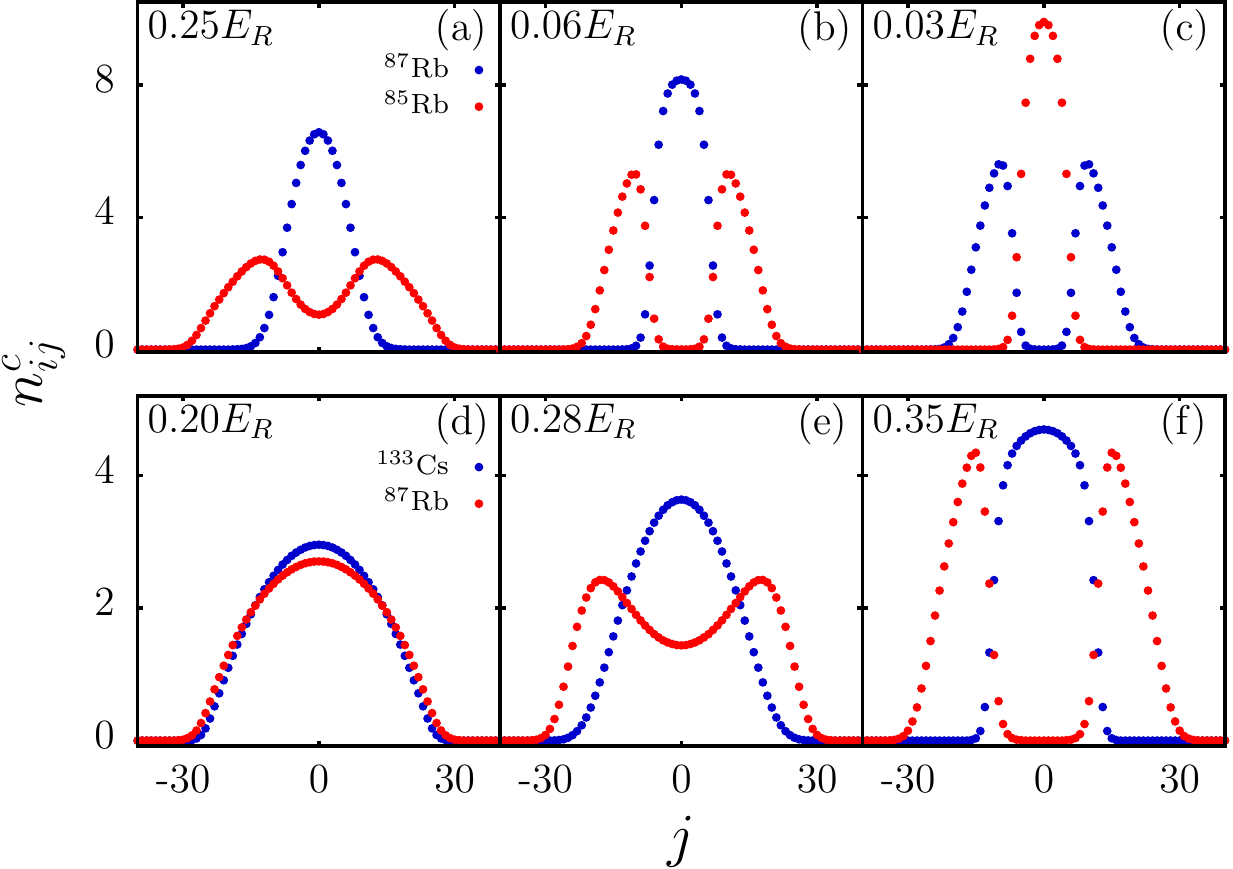}}
 \caption{The geometry of the condensate density profiles and its transition 
          from miscible to the immiscible regime. (a-c) The transition 
          from miscible to the \textit{sandwich} profile for 
          $^{87}$Rb-$^{85}$Rb TBEC with the change in the intraspecies 
          interaction $U_{22}$ at $T = 0K$. The position swapping (c) in the 
          \textit{sandwich} profile occurs at $U_{11} = U_{22} = 0.05 E_R$. 
          (d-f) Shows the similar condensate density profiles for Cs-Rb TBEC 
          with change in the interspecies interaction $U_{12}$ at $T = 0K$. 
          In this system the transition to \textit{sandwich} geometry occurs 
          at $U_{12}^c = 0.3 E_R$.}
 \label{den_2sp}
\end{figure}


\subsubsection{Third Goldstone mode in $^{87}$Rb - $^{85}$Rb TBEC}
To examine the mode evolution with the tuning of intraspecies interaction, 
we consider a quasi-1D TBEC consisting of  
$^{87}$Rb and $^{85}$Rb~\cite{papp_08,handel_11}. In this system, we 
consider $^{87}$Rb and $^{85}$Rb as the first and second species, 
respectively. The axial trapping frequency for both the species is 
$\omega_z = 2\pi\times 80$ Hz with the anisotropy parameters along $x$ and 
$y$ directions as $12.33$. The laser wavelength used to create the optical
lattice potential is $\lambda_L=775$ nm. The number of atoms are 
$N_{1} = N_{2} = 100$, which are confined in $100$ lattice sites superimposed 
on harmonic potential. We choose the depth of the lattice potential 
$V_0 = 5 E_R$ and set the tunneling matrix elements for the two species as 
$J_1 = 0.66 E_R$ and $J_2 = 0.71 E_R$, the intraspecies interaction 
$U_{11}$ as $ 0.05 E_R$ and the interspecies interaction $U_{12}$ as 
$0.1 E_R$. These set of DNLSE parameters are calculated by considering 
the width of the Gaussian beam as $0.3 a$. Since the scattering length of 
$^{85}$Rb is tunable with the Feshbach resonance~\cite{papp_08}, 
we study the excitation spectrum with the variation in $U_{22}$. The evolution 
of the Kohn mode functions with the variation of $U_{22}$ is shown in 
Fig.~\ref{mode_fns_775}. For $0.18 \le U_{22} \le 0.25 E_R$, the system is in 
the miscible domain, and the Kohn mode is a linear combination of $^{87}$Rb 
and $^{85}$Rb Kohn modes. As we approach the phase separation by reducing the 
value of $U_{22}$, we observe a decrease in the $^{87}$Rb component of the 
Kohn mode function amplitude and the mode component of $^{85}$Rb goes soft at 
$0.062 E_R$. The softening of the mode is evident from the evolution of the 
mode energies as shown in Fig.~\ref{mode_en}(a). The figure shows that the
mode continues as the third Goldstone mode for $U_{22} \le 0.062 E_R$. The 
emergence of the third Goldstone mode is associated with a change in the 
geometry of the system, the density changes from overlapping to 
\textit{sandwich} profile as shown in Figs.~\ref{den_2sp}(a-c). Thus, as 
discussed in our earlier work~\cite{arko_14}, the binary condensate is 
separated into three distinct sub-components.
\begin{figure}[ht]
 {\includegraphics[width=8.5cm] {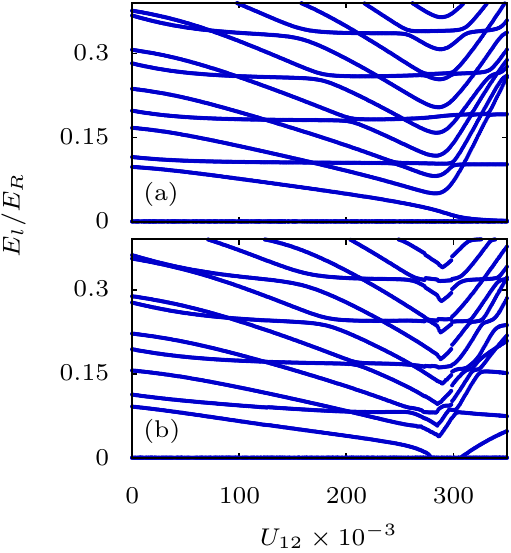}}
  \caption{The evolution of the energies of the low-lying modes as a function 
           of the interspecies interaction in Cs-Rb ($U_{12}$) TBEC held in 
           a quasi-1D lattice potential. (a) The excitation spectrum at 
           $T = 0K$, and (b) excitation spectrum after including the quantum
           fluctuations. Here $U_{12}$ is in units of the recoil energy $E_R$.}
  \label{mode_en_cs_rb}
\end{figure}


\subsubsection{Third Goldstone mode in $^{133}$Cs - $^{87}$Rb TBEC}
\begin{figure}[ht]
 {\includegraphics[width=8.5cm] {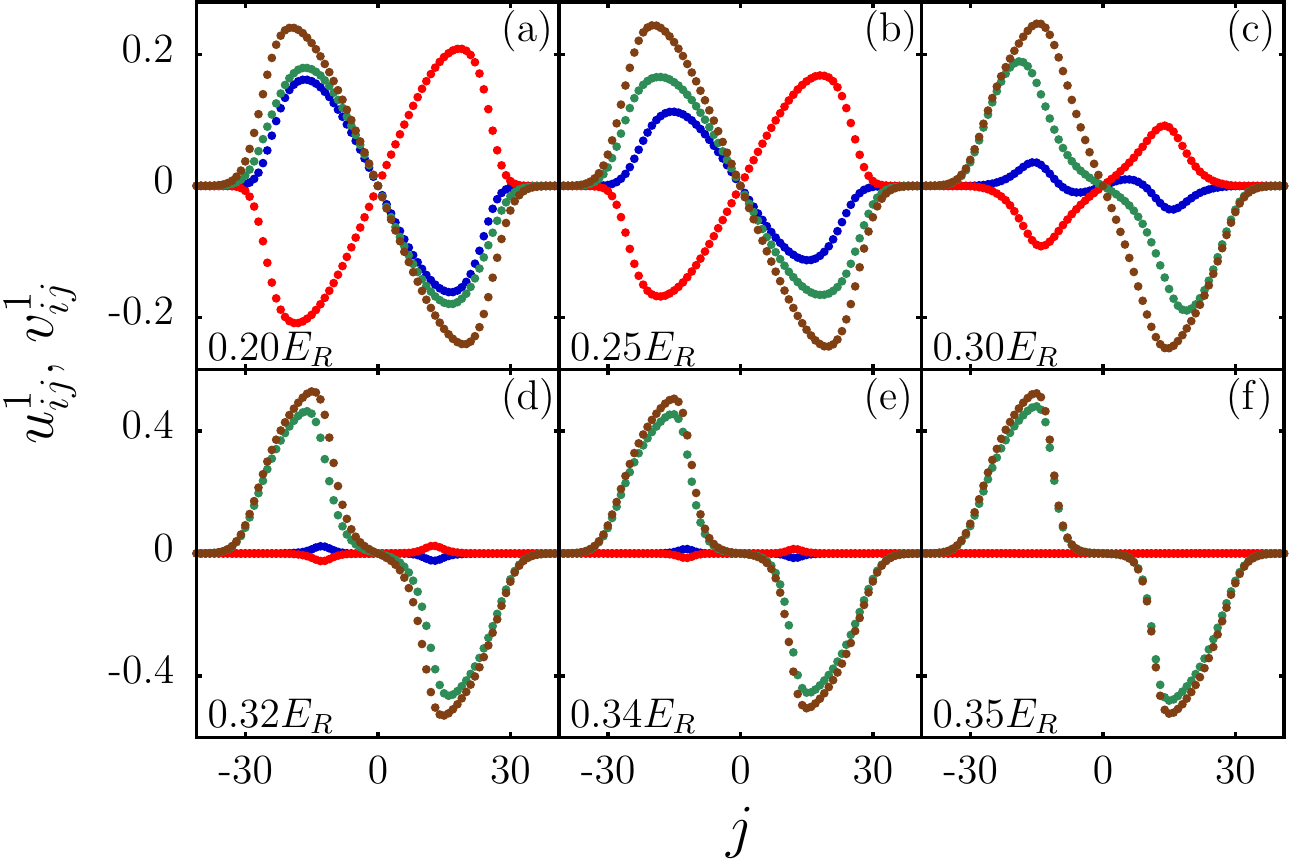}}
  \caption{The evolution of the quasiparticle amplitudes corresponding to
           the Kohn mode as the interspecies interaction is increased from  
           $0.2 E_R$ to $0.35 E_R$ for Cs-Rb TBEC in quasi-1D lattice 
           potential at $T = 0K$. (a-c) In miscible regime, the Kohn mode 
           has contributions from both the species. (d-f) For
           $U_{22}>0.3 E_R$ the Kohn mode of $^{87}$Rb goes soft, 
           whereas that of $^{133}$Cs decreases in amplitude.}
 \label{mode_fn_cs_rb}
\end{figure}

 For mode evolution with the tuning of interspecies interaction, we consider 
the binary system of Cs-Rb~\cite{mccarron_11,lercher_11}. Here, we consider 
$^{133}$Cs and $^{87}$Rb as the first and second species, respectively. To 
study the modes evolution as the system undergoes transition from miscible to 
immiscible phase, the interspecies interaction $U_{12}$ is varied, which is 
possible with magnetic Feshbach resonance~\cite{pilch_09}. The parameters of 
the system considered are $N_{1} = N_{2} = 100$ with the similar trapping 
frequencies as in the case of $^{87}$Rb-$^{85}$Rb mixture. The lattice 
parameters are chosen as $J_1 = 0.92 E_R, J_2  = 1.95 E_R, U_{11} = 0.40 E_R$, 
and $U_{22} = 0.21 E_R$. At $U_{12} = 0$, the two condensates are uncoupled 
and have two Goldstone modes, one corresponding to each of the two species. 
At low values of $U_{12}$, in the miscible regime, the condensate density
profile of both the species overlap as shown in Fig.~\ref{den_2sp}(d). As we 
increase $U_{12}$, the Kohn mode of $^{87}$Rb gradually goes soft and at a 
critical value $U_{12}^{c} = $$0.3 E_R$ it is transformed into the third
Goldstone mode. For $U_{12}^{c}<U_{12}$, the geometry of the condensate
density profile changes and acquires \textit{sandwich} structure in which the 
Cs condensate (higher mass) is at the center and flanked by Rb condensate
(lower mass) at the edges as shown in Fig.~\ref{den_2sp}(f). This is also 
evident from the evolution of the low-lying modes, shown in 
Fig.~\ref{mode_en_cs_rb}(a) and is reflected in the structural evolution of 
the quasiparticle amplitudes in Fig.~\ref{mode_fn_cs_rb}. Hence the system 
attains an extra Goldstone mode after transition from miscible to 
\textit{sandwich} type profile. 

\begin{figure}[ht]
 {\includegraphics[width=8.5cm] {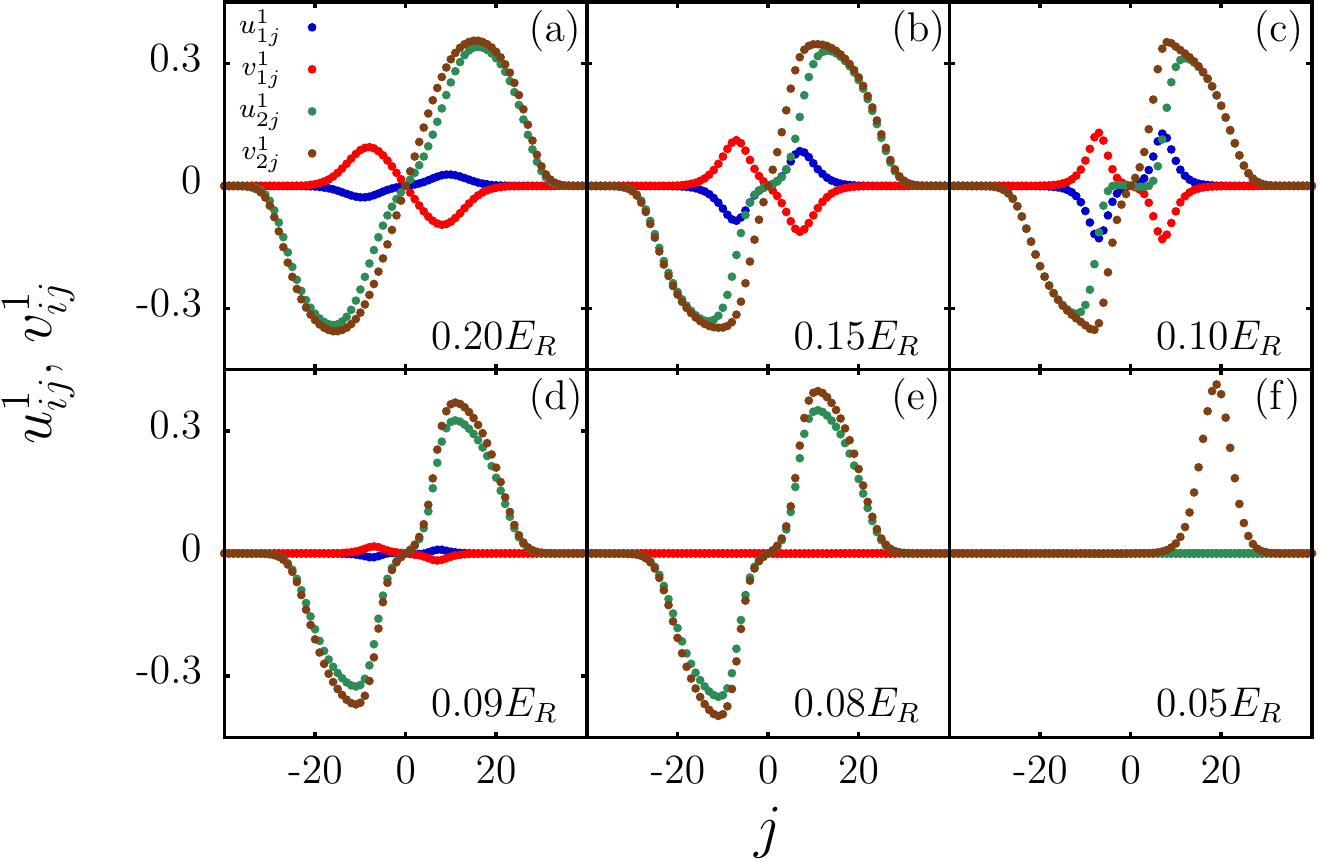}}
 \caption{The evolution of the quasiparticle amplitudes corresponding to 
          the Kohn mode for $^{87}$Rb-$^{85}$Rb TBEC in the presence of the 
          fluctuations as the intraspecies interaction of $^{85}$Rb ($U_{22}$) 
          is decreased from $0.2 E_R$ to $0.05 E_R$. (a-e) The Kohn mode of 
          $^{85}$Rb goes soft, whereas that of $^{87}$Rb is decreases in
          amplitude and finally vanishes in (e). (f) The sloshing mode, 
          which emerges after phase separation as the \textit{sandwich}
          density profile transforms into \textit{side-by-side} profile.}
 \label{mode_fns_fluc}
\end{figure}

\subsubsection{Position swapping of species}
 A remarkable feature in the evolution of the condensate density profiles of 
$^{87}$Rb-$^{85}$Rb TBEC with the variation of $U_{22}$ is the 
observation of the position swapping in the immiscible domain. This 
is absent when the trapping potential consists of only the harmonic 
potential (continuous system), and is the result of the discrete symmetry
associated with the optical lattice. 
As discussed earlier, in this system, we fix $U_{11}$ and $U_{12}$, and 
vary $U_{22}$ (intraspecies interaction of $^{85}$Rb). At higher values of 
$U_{22}$ the TBEC is in the miscible  phase, and as we decrease $U_{22}$, at
the critical value $U_{22}^{c} = 0.17 E_R$ the TBEC enters the immiscible 
domain. The geometry of the density profiles is \textit{sandwich} type and the 
component with smaller $U_{ii}$ is at the centre. An example of condensate
density profile in this domain, $U_{22}=0.06E_R$, is shown in 
Fig.~\ref{den_2sp}(b). In the figure, the species with smaller intraspecies 
interaction ($^{87}$Rb) is at the center and $^{85}$Rb is at the edges. As 
$U_{22}$ is further decreased, the system continues to be in the same phase. 
During evolution, an instability arises when both intraspecies interactions 
are same ($U_{11}=U_{22} = 0.05$). At this value of $U_{22}$ the components 
swap their places in the trap. This is also reflected in the excitation 
spectrum, a discontinuity at $U_{22} = 0.05 E_R$ in the plot of mode evolution 
shown in Fig.~\ref{mode_en}(a) is a signature of the instability. On further 
decrease of $U_{22}$, we enter the $U_{22}<U_{11}$ domain and $^{85}$Rb 
occupies the center of the trap.  An example of density profiles in this 
domain, $U_{22} = 0.03$ is shown in Fig.~\ref{den_2sp}(c). The position 
swapping, however, does not occur in Cs-Rb system as in that case we vary 
$U_{12}$.

\begin{figure}[h]
 {\includegraphics[width=8.5cm] {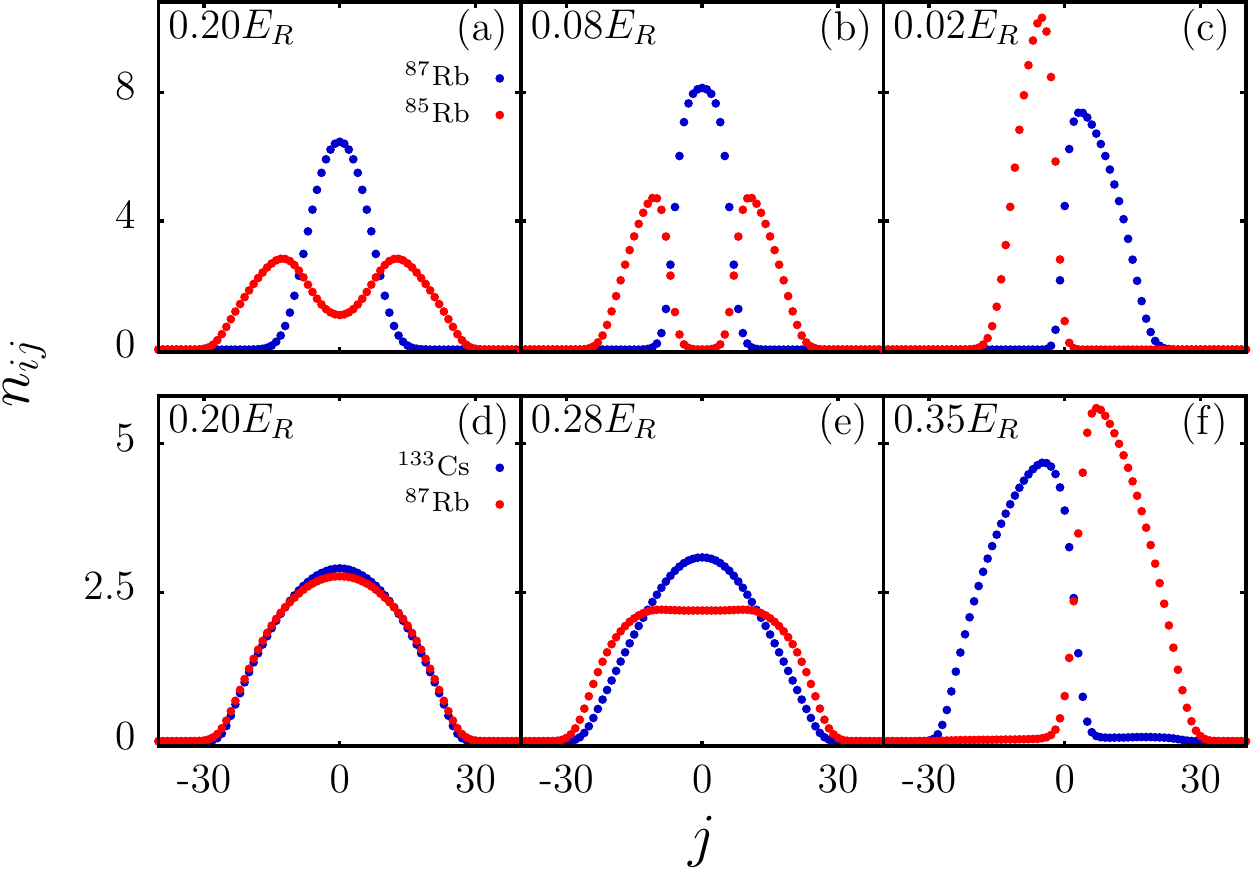}}
 \caption{The fluctuation induced transition in the geometry 
          of the total density profile (condensate + quantum fluctuations)
          of TBEC at $T = 0K$ in quasi-1D lattice potential. (a-c) The 
          transition in $^{87}$Rb-$^{85}$Rb system from miscible to
          \textit{sandwich} and finally in the \textit{side-by-side} profile 
          with the change in the intraspecies interaction. (d-e) The 
          transition in Cs-Rb TBEC from miscible to \textit{side-by-side} 
          profile with the change in interspecies interaction $U_{12}$. The 
          geometry of the ground state of both system in the immiscible regime 
          is different from that at zero temperature in the absence of the 
          fluctuations~Fig.\ref{den_2sp}.}
 \label{den_2sp_fluc}
\end{figure}

\begin{figure}[h]
 {\includegraphics[width=8.5cm] {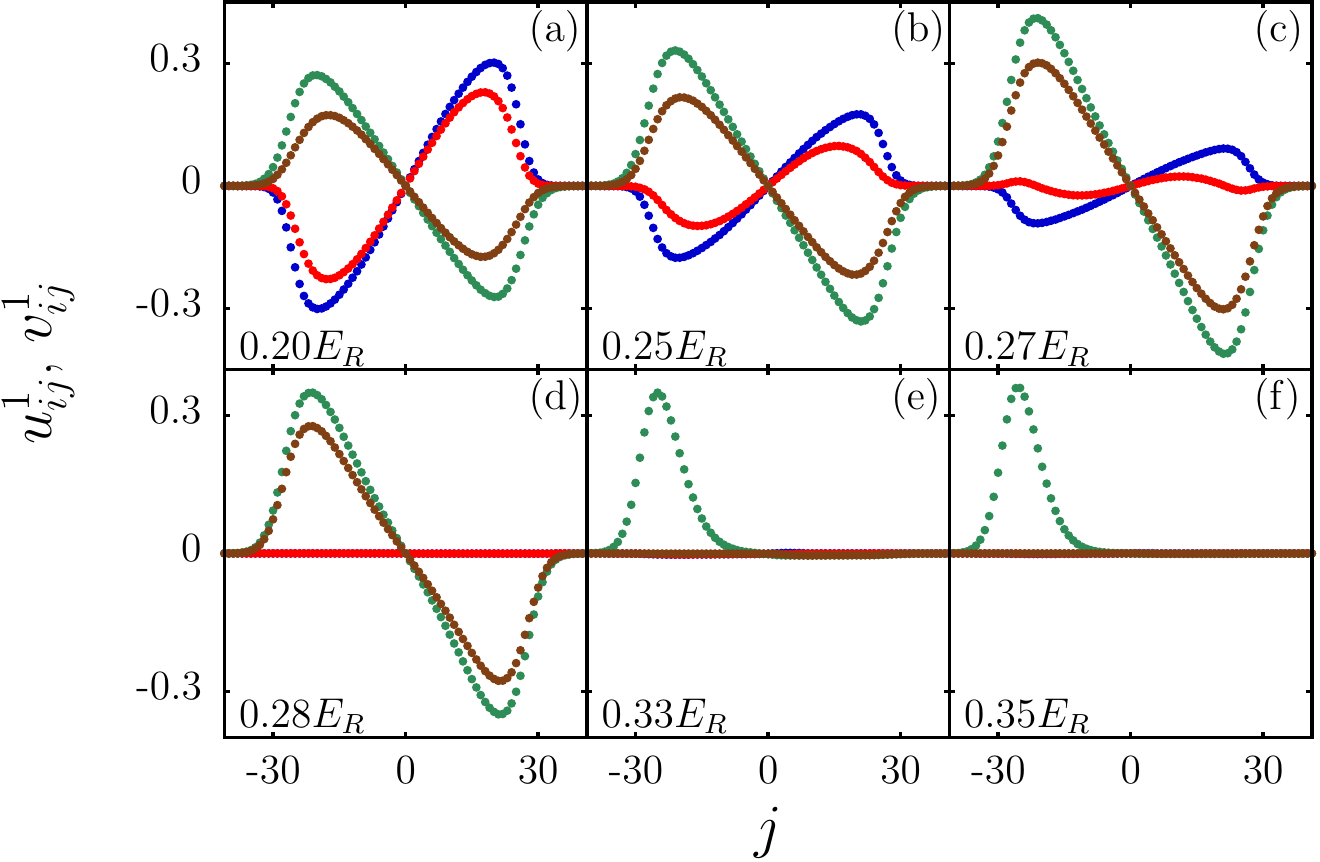}}
  \caption{The evolution of the quasiparticle amplitude corresponding to
           the Kohn mode for Cs-Rb TBEC in the presence of fluctuations as
           (a-d)The Kohn mode evolves as the interspecies interaction is 
           increased. (e-f) It is transformed into a sloshing mode as the 
           TBEC acquires \textit{side-by-side} density profile after phase 
           separation.}
  \label{mode_fn_cs_rb_fl}
\end{figure}


\subsection{Effect of quantum fluctuations}
\label{quan_fl}
 We compute the condensate profiles and modes for $^{87}$Rb-$^{85}$Rb TBEC,
however, include the effect of quantum fluctuations. We, then, encounter
severe oscillations in the number of atoms during iterations to solve the
DNLSEs and there is no convergence. To mitigate this, we use successive 
under-relaxation technique with $r^{\rm un} = 0.6$. For computations, we 
consider the same set of parameters as in the case of $T=0K$ without 
fluctuations. The fluctuations break the spatial symmetry of the system as we 
vary the intraspecies interaction of $^{85}$Rb ($U_{22}$). In the immiscible 
domain, the condensate density profile changes from \textit{sandwich} to 
\textit{side-by-side} profile at $0.078 E_R$. The system acquires a new stable 
ground state as the chemical potential of the system decreases from 
$0.92 E_R$ to $0.80 E_R$. The evolution of the mode energies with $U_{22}$ 
including the fluctuation is shown in Fig.~\ref{mode_en}(b). It is evident 
that at this value $U_{22} = 0.078 E_R$, the $^{85}$Rb Kohn mode goes soft and 
emerges as a sloshing mode. The transformations in the mode functions as 
$U_{22}$ is decreased about this point are shown in Fig.~\ref{mode_fns_fluc}. 
This topological phase transition is evident from the density profiles of the 
TBEC in the presence of the quantum fluctuations as shown in 
Fig.~\ref{den_2sp_fluc}(a-c).

In the Cs-Rb system, due to quantum fluctuations, the Kohn mode of $^{87}$Rb 
goes soft at a lower value of $U_{12}$ compared to the value without 
fluctuations. This is evident in the mode evolution with quantum fluctuations 
as shown in Fig.~\ref{mode_en_cs_rb}(b). The discontinuity in the spectrum is 
the signature of the transition from miscible to immiscible regime. The soft 
Kohn mode gains energy and gets hard at $0.31 E_R$. This mode hardening is 
due to the topological change in the ground state density profile from 
miscible to the \textit{side-by-side} profile, shown in 
Fig.~\ref{den_2sp_fluc}(d-f). The lowest mode with nonzero excitation energy 
corresponding to the \textit{side-by-side} profile is shown in 
Fig.~\ref{mode_fn_cs_rb_fl}(e-f).


\section{Conclusions}
\label{conc}
 We have studied the ground state density profiles and the excitation spectrum 
of TBEC in quasi-1D optical lattices. We observe that the system gains an 
additional Goldstone mode at phase-separation at zero temperature. Furthermore, in TBEC where miscible to immiscible transition driven through the variation of the intraspecies interaction ($^{87}$Rb-$^{85}$Rb), a finite discontinuity in 
the excitation energy spectra is observed in the neighbourhood of equal 
intraspecies interaction strengths. In the presence of quantum fluctuations, 
on varying the intraspecies interaction of $^{85}$Rb, in the immiscible 
regime, the ground state density profiles transform from \textit{sandwich} to 
\textit{sie-by-side} geometry. This is characterized by the hardening of the 
Kohn mode which emerges as a sloshing mode. The fluctuation induced topological 
change from completely miscible to \textit{side-by-side} ground state density 
profile is also evident in $^{133}$Cs-$^{87}$Rb mixture. Our current studies 
show that the geometry of the density profiles with and without quantum 
fluctuations are different. Since quantum fluctuations are present in 
experiments, it is crucial to include quantum fluctuations to obtain correct 
density profiles of TBECs in optical lattices in the phase-separated domain.

\begin{acknowledgments}
 We thank S. Gautam and S. Chattopadhyay for useful discussions. The results 
presented in the paper are based on the computations using the 3TFLOP HPC 
Cluster at Physical Research Laboratory, Ahmedabad, India. 
\end{acknowledgments}

\bibliography{tbec_opl_mode}{}
\bibliographystyle{apsrev4-1}
\end{document}